\documentclass[aps,prapplied,preprint,12pt,groupedaddress,notitlepage]{revtex4-2}  
\usepackage[utf8]{inputenc}
\usepackage[T1]{fontenc}
\usepackage[english]{babel}
\usepackage{amsmath,amssymb,graphicx,wrapfig,setspace,placeins,csquotes,booktabs}
\begin{document}
\preprint{APS/123-QED}
\title{The Thermal Cost of Harvesting the Solar Infrared Tail in Space-Based Photovoltaics}

\author{David E. Abraham}
\author{Linus Kim}
\author{Aaswath P. Raman}
\email{aaswath@ucla.edu}

\affiliation{Department of Materials Science and Engineering, University of California, Los Angeles, Los Angeles, CA 90095, USA}

\begin{abstract}

Space applications require the highest performing photovoltaic devices. Typically these are multi-junction devices that are capable of converting a larger bandwidth of the solar spectrum. However, the conversion of longer wavelength photons in the solar infrared tail results in a disproportionate increase in generated heat per watt of electricity, and since devices have limited heat rejection in space, this excess heat raises the operating temperature, lowering the efficiency of the entire device. Here we show that space-based photovoltaics should avoid converting the solar infrared tail past an optimum terminal cutoff wavelength, and that this will lead to higher electrical output. In addition, a lower operating temperature may lead to higher end-of-life power outputs.   

\end{abstract}

\maketitle

\section{Introduction}

Since Shockley and Queisser published their detailed balance limit on the conversion efficiency of a single-junction solar cell, \cite{shockley_detailed_1961}, there has been a concerted effort to develop multi-junction devices with higher efficiencies. Three- and four-junction III-V devices with efficiencies around 32\% are commercially available and are widely used in space applications where ultra-high performance is essential.\cite{azurspace2019-4g32c-advanced,azurspace2024-3g30c,spectrolab2023xte447} To push efficiencies even higher, recent work has sought to increase the number of junctions to broaden the absorption band and harvest more of the solar infrared (IR) tail. The current record research-grade device under a single sun is a six-junction inverted metamorphic cell developed by NREL with a terminal cutoff wavelength, $\lambda_{cutoff}$, of 1800~nm and an efficiency of 39.2\% under AM1.5G at room temperature. \cite{geisz_six-junction_2020}

However satellites in orbit often see operating temperatures much higher than room temperature, with those in low‐Earth‐orbit (LEO) typically reaching temperatures as high as 80~\textdegree C \cite{ali_design_2018, landis_review_1994}. In contrast, measurements made on the Planck satellite reveal solar cell temperatures ranging from 104-125~\textdegree C with a mean of 111~\textdegree C (the satellite orbits around the Sun-Earth $L_2$ point roughly $1\%$ farther from the sun than a satellite in LEO)\cite{planck_collaboration_planck_2011}. 

These higher operating temperatures critically influence a solar cell's conversion efficiency. III–V-material-based solar cells see an efficiency drop of roughly $0.05-0.10~\%$ absolute per Kelvin, and the temperature coefficient increases as the device sustains radiation damage over its lifetime \cite{siefer_analysis_2014, azurspace2019-4g32c-advanced,azurspace2024-3g30c,spectrolab2023xte447}. Roughly 80\%-90\% of this loss is caused by a decrease in $V_{OC}$, stemming from increased recombination due to increased carrier concentrations \cite{green_general_2003}. Thus minimizing the operating temperature is crucial to realize the high-performance of multi-junction solar cells.

But reducing the operating temperature of photovoltaics in space is challenging. The only viable heat rejection pathway in space is radiative cooling, governed by the Stefan–Boltzmann law,\cite{Stefan1879,Boltzmann1884} given by $Q = \varepsilon \sigma T^4$ where $Q~[W/m^2]$ is the radiated power, $\varepsilon$ is the total hemispherical emissivity and $\sigma=5.67\cdot10^{-8}~[W/m^2/K^4]$ is the Stefan-Boltzmann constant. For a solar cell array at a given temperature, the only way to increase its cooling power is thus to either increase the area available for emission\cite{cannon_internally_2025} or increase the emissivity of the radiating surfaces, both of which have been proposed for both the active \cite{li_comprehensive_2017} and passive \cite{shimizu_blackbody_2014} surfaces. Commercial photovoltaics for space have minimum emissivities of around 0.85 \cite{qioptiq-coverglasses}. Ultimately, however, radiative cooling fundmentally constrains thermal rejection. \emph{Reducing} heat generation is thus an indispensable avenue to reducing the operating temperature. 

Here we demonstrate a strategy which allows space-based photovoltaics to operate at lower temperatures by intentionally not converting the solar infrared tail. In particular we elucidate the thermal cost of harvesting the solar infrared tail in space-based photovoltaics given fundamental limits on radiative heat transfer from a solar module. We analyze how longer‐wavelength photons generate increasingly more heat per watt of electrical power than shorter‐wavelength photons. We ultimately show that, under AM0 illumination and realistic radiative cooling constraints, space‐based multi‐junction cells can generate more electrical power by truncating their conversion band before the room-temperature optimum point. This in turn suggest that surprisingly different multi-junction designs may prove more optimal when operating in space-based environments.

\section{Results}

\subsection{Radiative Constraints and Efficiency Limits}

\begin{figure}[hbt]
\centering
\includegraphics[width=0.9\linewidth]{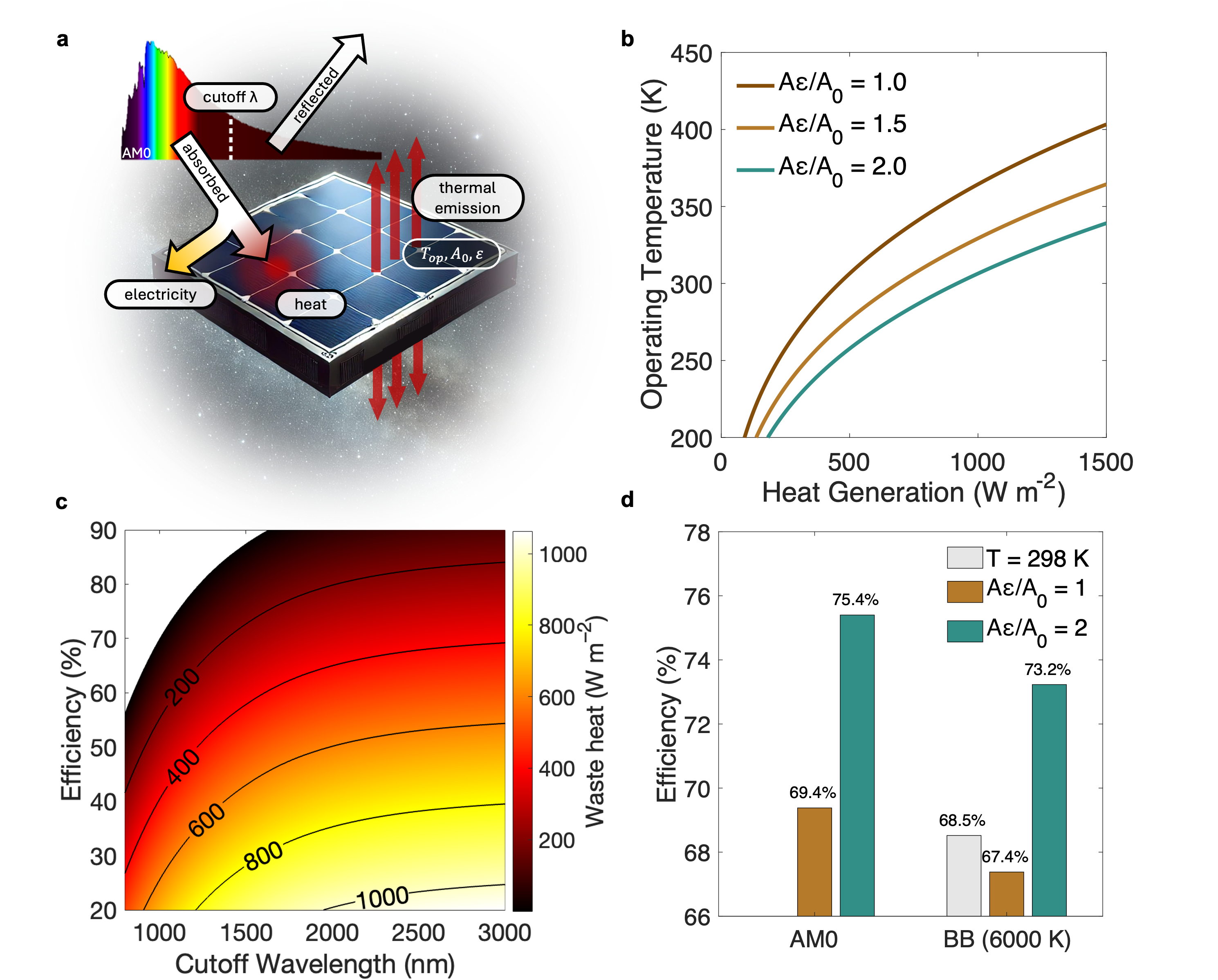}
\caption{\textbf{Effect of radiative cooling constraints on the operating temperature and theoretical efficiency limits of ideal multijunction solar cells in space.} (a) Schematic of a solar cell operating in space under AM0 illumination, with perfect sub‐bandgap photon reflection and constrained waste‐heat removal (combined front and rear emissivity limited to $\epsilon = 2$). (b) Operating temperature of the cell as a function of waste‐heat flux for effective radiating factors $A\epsilon/A_0 = 1.0$, $1.5$, and $2.0$. (c) Waste heat generation plotted as a function of the cell’s cutoff wavelength and nominal conversion efficiency. (d) Detailed-balance efficiency limit for an infinite-junction solar cell at $T=298\,$K, under AM0 and a $6000\,$K blackbody spectrum. Bars compare De Vos's reference result at $T=298\,$K (grey) with predicted efficiencies for effective radiating factors $A\epsilon/A_0 = 1.0$ (brown) and $2.0$ (teal). Under AM0 illumination the limits are $69.4\%$ ($A\epsilon/A_0 = 1.0$) and $75.4\%$ ($A\epsilon/A_0 = 2.0$); under $6000\,$K blackbody they are $67.4\%$ and $73.2\%$, versus De Vos's $68.5\%$ \cite{vos_detailed_1980}
}
\label{fig:figure1}
\end{figure}

Radiative cooling, whether active or passive, remains the only heat rejection mode available to satellites in space. The rate of heat rejection from a given surface is defined by the Stefan–Boltzmann law ($Q = A\,\epsilon\,\sigma\,T^4$). The satellite's operating temperature, $T$, is the temperature where the rate of heat rejection equals the rate of heat generation. In this case, the term $A\,\epsilon$ is an effective radiating factor capturing the spectral and spatial average emissivity of the satellite, and it has a maximum value of $2\,A_0$ for a two-sided blackbody which has perfect emissivity across both hemispheres. Thus, if the effective radiating factor and rate of heat generation is known, the operating temperature can be determined. Figure \ref{fig:figure1}b plots the operating temperature of a bifacial object in space with different effective radiating factors as a function of heat generation.

Shifting our attention to the photovoltaic array - the rate of heat generation can be determined if the photovoltaic conversion efficiency and the width of the spectral conversion band is known. The conversion band lies between the UV-cutoff, often dictated by the optical coating or packaging, and the smallest bandgap of a multi-junction solar cell. For simplificty, we assume that the device reflects all photons outside of this conversion band, and perfectly absorbs all photons within the conversion band. The heat generated is then given by:

\begin{align}
  P_{heat} = P_{AM0}-P_{electric}-P_{reflected} \\
  P_{heat} = P_{AM0}\,(1-\eta)-P_{reflected}
\end{align}

where $\eta$ is the nominal efficiency of the solar cell, defined as the electrical power generated divided by the total incident power (including photons outside of the conversion band), typically at room temperature. Figure \ref{fig:figure1}c plots the heat generated against the nominal efficiency and cutoff-wavelength of a solar cell, treating them as two independent variables even though in reality they are related. It is clear that for a given nominal efficiency, increasing the conversion band increases the amount of heat generation since the amount of absorbed power increases and a fraction of this absorbed power invariably becomes heat.

For a non-current matched photovoltaic with infinite junctions and electroluminescent coupling, the 1-sun efficiency is roughly 68\%.\cite{vos_detailed_1980} In their original analysis, De Vos used a blackbody at 6000~K to approximate the solar spectrum and sets the device temperature to 300~K. Here we present a slightly modified ideal limit for an infinite-junction tandem cell operating in space with the aforementioned radiative constraints and accounting for the real AM0 spectrum. The efficiency and terminal cutoff wavelength of the device determine the waste heat generation which in turn determines the operating temperature; the operating temperature affects the conversion efficiency. We iteratively solve for the operating temperature and efficiency of an infinite-junction, radiative-limit tandem solar cell under AM0 and a 6000~K blackbody for direct comparison to De Vos (Figure \ref{fig:figure1}d). With an effective radiating factor of $2\,A_0$, De Vos' 68.5\% 6000~K blackbody efficiency is raised to 73.2\% as the operating temperature lowers to 239~K. In contrast, with an effective radiating factor of $1\,A_0$, the efficiency decreases to 67.4\%. The values for AM0 are shown as well. In neither case do we account for the earth's thermal emission, some fraction of which is absorbed by the cell and causes an increase in operating temperature and associated decrease in efficiency. 

\subsection{Ideal Multi-junction Solar Cells Operating in Space}

\begin{figure}[hbt]
\centering
\includegraphics[width=0.9\linewidth]{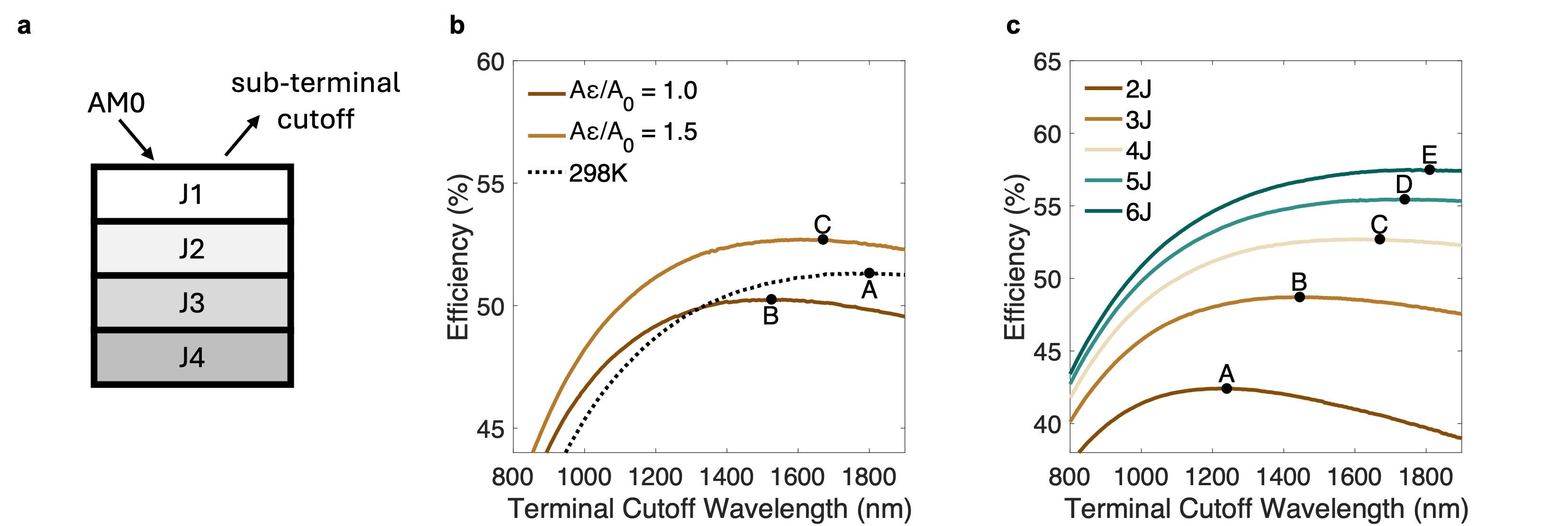}
\caption{\textbf{The effect of constrained thermal management on the optimum $\lambda_{cutoff}$ of ideal multijunction solar cells operating under AM0.} (a) Schematic of an ideal four-junction solar cell (with bandgaps $E_{g1}>E_{g2}>E_{g3}>E_{g4}$) operating at the detailed-balance efficiency limit. The device perfectly reflects photons with energy below the terminal cutoff. (b) AM0 conversion efficiency as a function of $\lambda_{cutoff}$ for a four-junction cell with constrained thermal-radiation conditions (solid lines) and for the same cell held at $298\,$K (dashed line). Limited heat rejection raises the cell temperature at longer cutoff wavelengths causing the peaks to shift to shorter cutoff wavelengths. Point A denotes the $298\,$K optimum at $\lambda=1800\,$nm ($\eta=51.3\%$), B marks the $A\varepsilon/A_{0}=1.0$ optimum at $\lambda=1525\,$nm and $T=308\,$K ($\eta=50.3\%$), and C the $A\varepsilon/A_{0}=1.5$ optimum at $\lambda=1670\,$nm and $T=279\,$K ($\eta=52.7\%$). (c) Operating AM0 efficiency versus $\lambda_{cutoff}$ for ideal $N$-junction ($N=2$--6) solar cells (radiative-recombination-limited) with $A\varepsilon/A_{0}=1.5$. Point A: $42.4\%$, 1240\,nm ($T=279\,$K); Point B: $48.7\%$, 1445\,nm ($T=278\,$K); Point C: $52.7\%$, 1670\,nm ($T=279\,$K); Point D: $55.5\%$, 1740\,nm ($T=275\,$K); Point E: $57.5\%$, 1810\,nm ($T=273\,$K).}
\label{fig:figure2}
\end{figure}

We next show that the optimum terminal cutoff of an ideal multi-junction solar cell operating in space is different from its room temperature, nominal efficiency. Figure \ref{fig:figure2}a shows a finite-junction solar cell operating under AM0 with perfect reflections of photons with energy below the bandgap of the smallest junction. For a given number of junctions, there exists an optimum terminal cutoff. We implement a derivative of the detailed balance, Shockely Quiesser limit which is detailed in Methods to current match the junctions, given $\lambda_{cutoff}$. We can then determine the optimum $\lambda_{cutoff}$, where the efficiency of the device with current-matched junctions is maximized. In this section, we consider only radiative recombination, which represents the ideal case and produces the highest possible efficiency. The dotted line in Figure \ref{fig:figure2}b shows the optimum point, A, for the 4-junction radiative limit cell at 298 K. 

With this model, we are able to determine the heat generated by the current-matched device for each terminal-cutoff wavelength. By assigning an effective radiating factor, we can then iteratively determine the operating temperature. The calculation is iterative because temperature affects conversion efficiency, altering generated heat, which in turn affects the temperature. In Figure \ref{fig:figure2}b we show the efficiency as a function of $\lambda_{cutoff}$ for two effective radiating factors. Note that, unlike for the dotted line at 298 K, the operating temperature of the devices for the dark and light brown curves is not constant as the terminal cutoff changes. Figure \ref{fig:figure2}b shows that lower effective radiating factors result in a shorter optimum $\lambda_{cutoff}$, and that the penalty for extending to longer wavelengths is greater than from the constant temperature optimization (the slope of the curve is greater). Figure \ref{fig:figure2}c shows the optimum terminal cutoff for 2- through 6J solar cells, operating in space under AM0, with effective radiating factors of 1.5. As the number of junctions increases, more of the solar spectrum may be converted while keeping thermalization losses low, thus minimizing heat generation and keeping the operating temperature low. The curve with point C corresponds to point C in Figure \ref{fig:figure2}b. 

We emphasize that we are not merely referring to the change in efficiency resulting from a deviation from the optimum terminal cutoff for a given number of junctions. It is well known that for a given number of junctions and incident spectrum, the power generation is maximum at an optimum $\lambda_{cutoff}$ - above which (longer cutoffs) there is an increase in thermalization losses and a lowering of the open-circuit voltage, and below which (shorter cutoffs) there is an increase in transmission losses and a lowering of the short-circuit current. Instead, our calculation considers the temperature dependent performance, given the specific radiative constraints of a photovoltaic module in space, to determine the optimum terminal cutoff.

\subsection{Influence of Non-idealities}

\begin{figure}[hbt]
\centering
\includegraphics[width=0.9\linewidth]{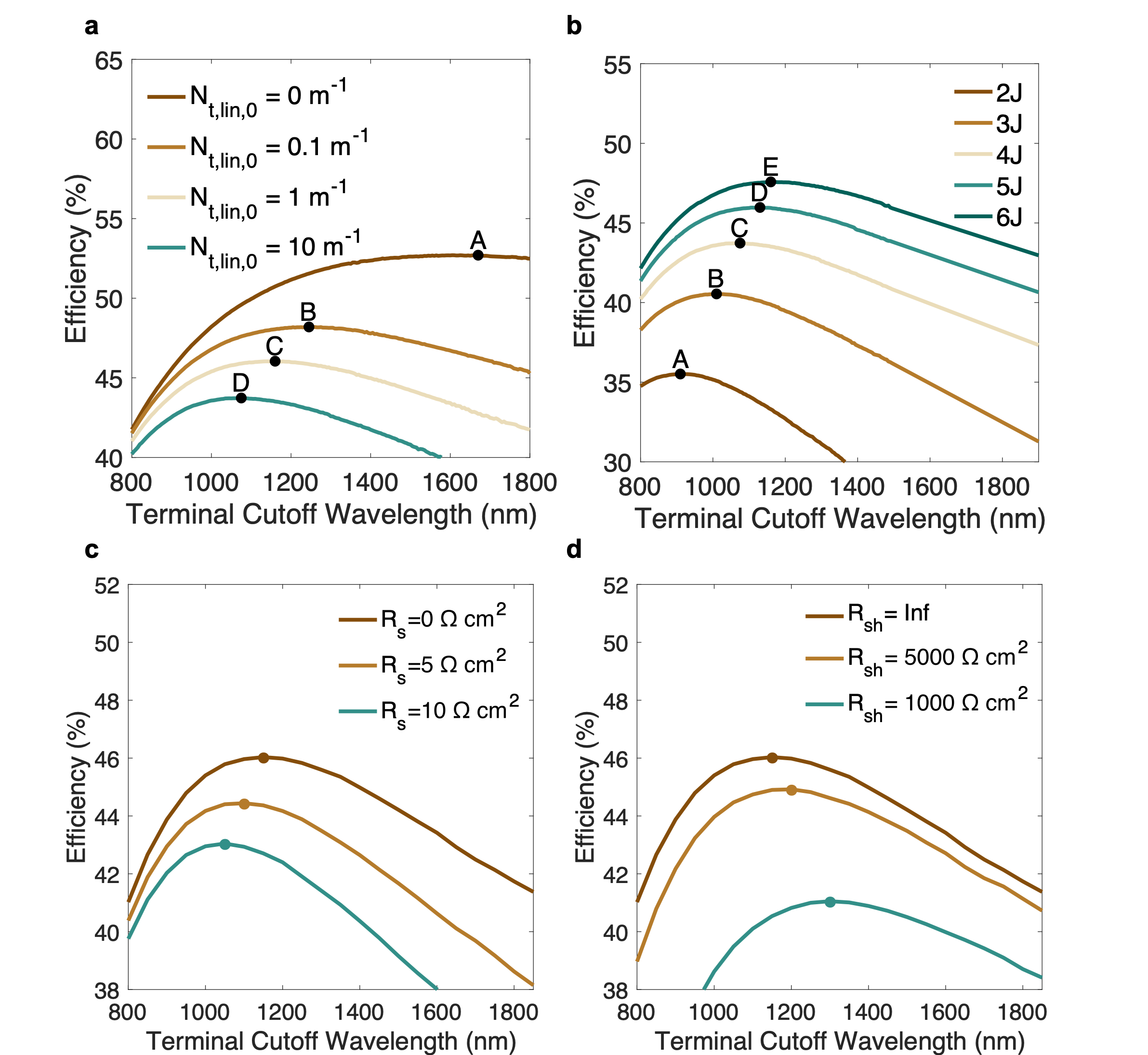}
\caption{\textbf{Effect of non‐radiative recombination, series resistance, and shunt resistance on optimum $\lambda_{cutoff}$.} (a) Shift in optimum cutoff wavelength for a four-junction cell under AM0 as non-radiative recombination increases ($A\varepsilon/A_{0}=1.5$). Point A shows the cutoff under the radiative-recombination limit at 1670\,nm ($\eta=52.7\%$), and points B–D show the cutoff shift to shorter wavelengths and the efficiency drop (Point B: $48.2\%$, 1245\,nm; Point C: $46.0\%$, 1160\,nm; Point D: $43.7\%$, 1075\,nm). (b) Optimum cutoff wavelengths for 2–6 junction cells with $A\varepsilon/A_{0}=1.5$ and $N_{t,\mathrm{lin},0}=1\ \mathrm{m}^{-1}$. Point A: $35.5\%$, 910\,nm ($T=261\,$K); Point B: $40.5\%$, 1010\,nm ($T=263\,$K); Point C: $43.7\%$, 1075\,nm ($T=263\,$K); Point D: $46.0\%$, 1130\,nm ($T=263\,$K); Point E: $47.6\%$, 1160\,nm ($T=262\,$K). (c) The effect of increasing series resistance with infinite shunt resistance for a four-junction cell ($A\varepsilon/A_{0}=1.5$), showing the efficiency drop and the optimum cutoff shift to shorter wavelengths. (d) The effect of decreasing shunt resistance with no series resistance, showing the efficiency drop and the cutoff move to longer wavelengths.}
\label{fig:figure3}
\end{figure}

Next, we introduce non-radiative recombination, which in general acts to decrease the efficiency and raise the operating temperature of the photovoltaic cell. We expect that this in turn should result in a shorter optimal $\lambda_{cutoff}$ compared to radiative-limit ideal cells previously analyzed. Here, non-radiative recombination groups together bulk, surface, and interface trap-assisted recombination; it does not include Auger recombination, which scales with the cube of carrier concentration and, in the case of III-V direct-bandgap solar cells, is insignificant even under concentration.\cite{vossier_is_2010} We model non-radiative recombination by determining a linear trap density which is a function of material bandgap and temperature (see Methods). 

Figure \ref{fig:figure3}a shows the effect of increasing non-radiative recombination from the ideal case, represented by the dark brown curve where the linear trap density equals 0. Increasing non-radiative recombination decreases conversion efficiency and causes the optimum terminal cutoff to shift to shorter wavelengths. In Figure \ref{fig:figure3}b, the optimum terminal cutoff as a funciton of the number of junctions is shown for $A\varepsilon/A_{0}=1.5$ and $N_{t,\mathrm{lin},0}=1\ \mathrm{m}^{-1}$. Once again, increasing the number of junctions decreases thermalization losses and associated heat generation, making it favorable to convert more of the solar infrared tail. 

Figure \ref{fig:figure3}a shows the efficiency of a 4-junction solar cell plotted against the $\lambda_{cutoff}$. The junction bandgaps for the top three junctions are optimized for AM0 given $\lambda_{cutoff}$. Two things are notable: first, the optimum terminal cutoff shifts to shorter wavelengths under the radiative emission constraints of space (for $A\,\epsilon=1.5$). Secondly, for long enough terminal-cutoff values, the operating efficiency actually decreases compared to the nominal efficiency, indicating that it is disadvantageous to convert that part of the spectrum. We plot the operating efficiency as a function of $\lambda_{cutoff}$ for 2- through 6- junction tandem solar cells. 

Increasing series resistance results in ohmic losses which, in addition to directly reducing the efficiency, also raise the operating temperature and result in temperature-driven efficiency losses. Figures \ref{fig:figure3}c and \ref{fig:figure3}d show the effect of increasing device series and shunt resistance.  As a result, the optimum terminal cutoff shifts to shorter wavelengths. In contrast, decreasing shunt resistance (increasing the number of shunts), causes the cutoff to move to longer wavelengths.

\subsection{Quantifying the Thermal Penalty of Extended IR Absorption}
Using a 100 junction cell, we calculate the efficiency versus terminal cutoff for both a radiative-limit ideal cell and a non-radiative recombination limited cell, with $N_{t,\mathrm{lin},0}=1\ \mathrm{m}^{-1}$. In both cases, the effective radiating factor is $A\varepsilon/A_{0}=1.5$ and the cells are operating under AM0. For each incremental increase in cutoff wavelength, we find the increase in generated heat, $dP_{heat}$, and electric power, $dP_{elec}$. The ratio of these two values shows the thermal cost per watt of electric power associated with converting that  

\begin{figure}[hbt]
\centering
\includegraphics[width=0.9\linewidth]{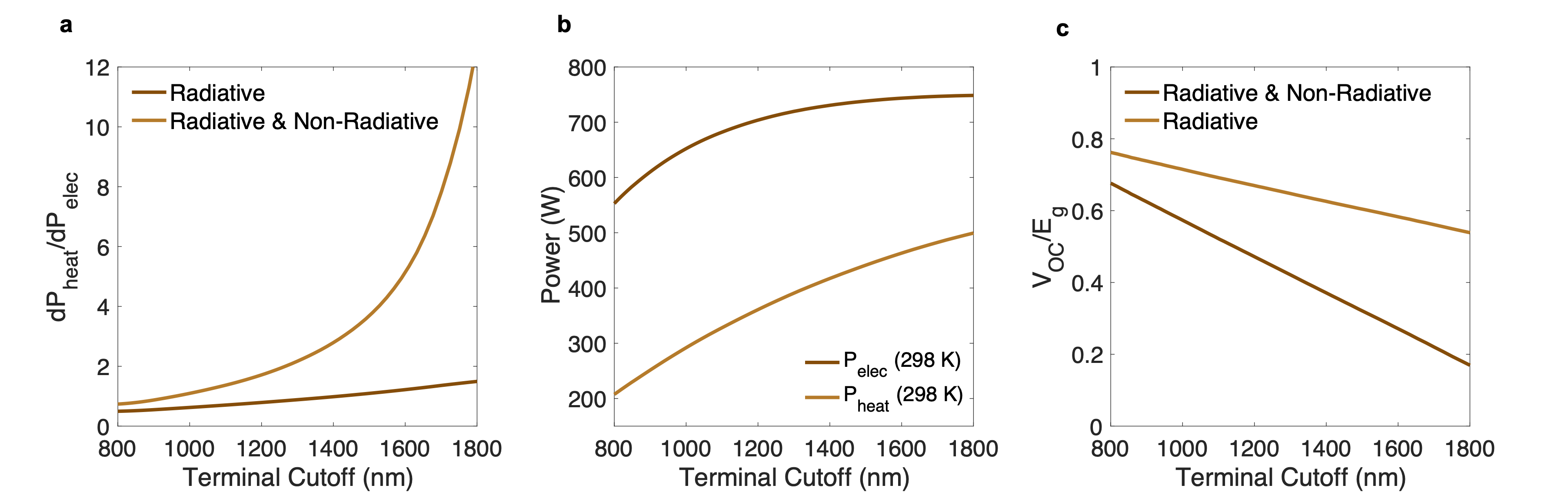}
\caption{\textbf{Deleterious increase in waste heat per watt of electricity as $\lambda_{cutoff}$ is extended.} (a) Spectral ratio of incremental thermal to electrical power plotted as a function of $\lambda_{cutoff}$ for radiative‐only (brown) and combined radiative + non‐radiative (orange) recombination conditions (AM0 illumination, $A\varepsilon/A_{0}=1.5$, $R_s=0$, and $R_{sh}=\,$Inf). The rising trend with increasing cutoff wavelength reveals that longer‐wavelength cutoffs produce progressively more heat per watt of electricity generated. (b) Electrical (dark brown) and thermal (light brown) power output versus $\lambda_{cutoff}$ for a solar cell with both radiative and non‐radiative recombination at $T=298\,$K. As the cutoff is shifted to longer wavelengths, the electrical power curve levels off, while the thermal power continues to rise—elucidating the increasing $dP_{\mathrm{heat}}/dP_{\mathrm{elec}}$ trend seen in \ref{fig:figure4}a. (c) Ratio of the open circuit voltage to the bandgap of the smallest bandgap junction, showing the decrease in useful work for lower energy phtons.}
\label{fig:figure4}
\end{figure}

In order to find the incremental increase in heat per increase in electric power, we start with a large number of junctions and add current matched junctions. This ensures that the performance of previous junctions is constant as more junctions are added and so that we are then only seeing the effect of the additional junction. We begin with 50-junctions for a cutoff wavelength of 800~nm, and by the time the terminal cutoff is 2000~nm, there are 141 junctions. Throughout this process, we use a high resolution spectrum ($d\lambda = 0.001~nm$). Figure \ref{fig:figure4}a plots $dP_{\mathrm{heat}}/dP_{\mathrm{elec}}$ as a function of terminal cutoff for both a radiative and non-radiative recombination limited cells. In both cases, there is a disproportionate rise of heat generation versus electrical output. For the non-radiative case, this increase is dramatic. Figure \ref{fig:figure4}b elucidates that the electrical power generated plateaus sooner than the heat power generated, thus increasing the thermal cost per watt. 

Finally, in \ref{fig:figure5}c, we plot the fraction of bandgap energy that is open-circuit voltage, which decreases as $\lambda_{cutoff}$ increases. The open circuit voltage is the difference betwen the bandgap and the sum of the fermi-level offsets for the n- and p-type materials. These offsets are largely set by the doping concentration and temperature, and do not change proportionally to the bandgap. Thus, as the bandgap decreases while these offsets stay constant, the relative amount of useful work decreases, using $V_{oc}$ as a proxy.

\subsection{Realistic Space Photovoltaic Performance}

\begin{figure}[hbt]
\centering
\includegraphics[width=0.9\linewidth]{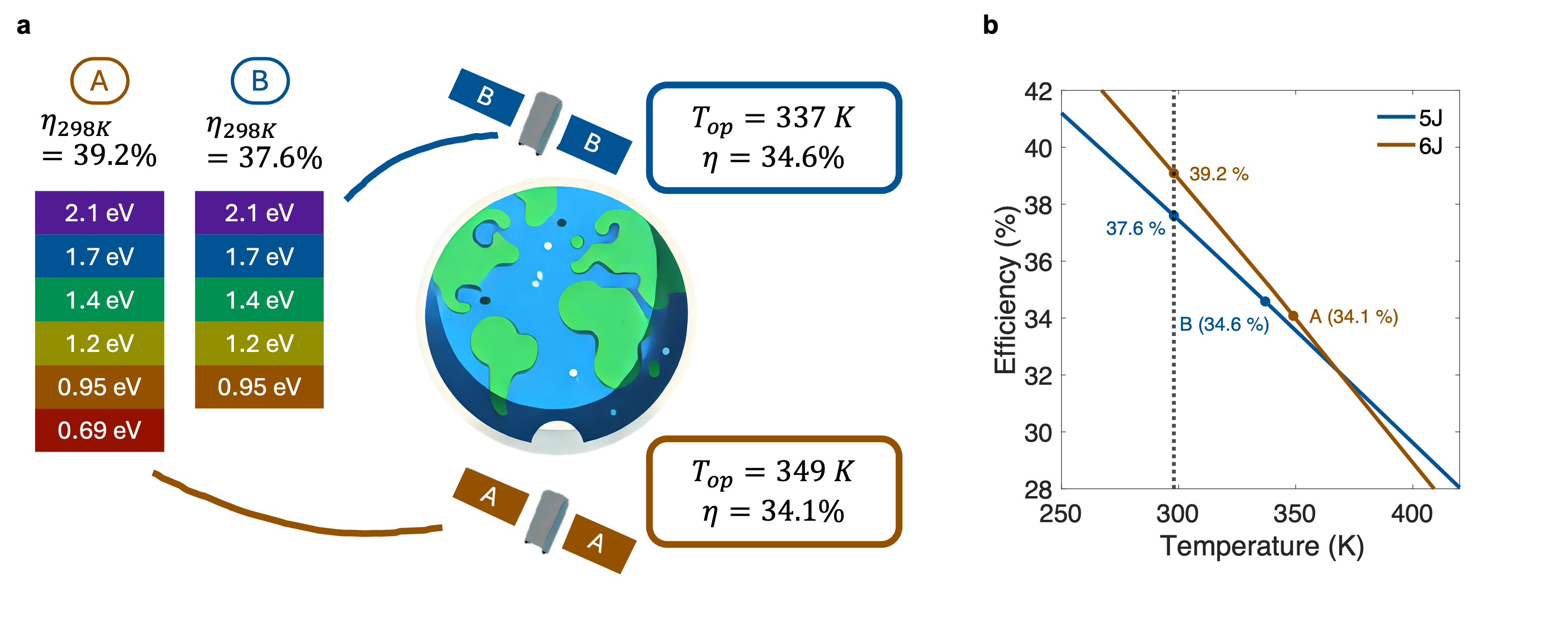}
\caption{\textbf{Temperature‐dependent performance of NREL 6‐junction and truncated 5‐junction solar cells under AM1.5G illumination.} The baseline 6‐junction cell (cell A; nominal $\eta=39.2\%$ under AM1.5G) and a variant with its bottom junction removed (cell B; nominal $\eta=37.6\%$ under AM1.5G, cutoff shifted from 1800~nm to 1300~nm) are both modeled including non‐radiative recombination and thermal radiation from Earth (approximated as a $255\,$K blackbody). In space, cell A heats to $71\,^\circ\mathrm{C}$ and its efficiency drops to $\eta=34.1\%$, while cell B reaches only $39\,^\circ\mathrm{C}$ and achieves a higher in‐space efficiency of $\eta=34.6\%$. (b) Temperature dependence of the efficiency of the baseline 6J device and the truncated 5J device. The broken line dictates room temperature, where the 6J cell outperforms the 5J cell. In contrast, points A and B mark the operating efficiency of each cell in orbit, where the 5J device outperforms the 6J despite converting less of the solar spectrum. 
}
\label{fig:figure5}
\end{figure}

We illustrate this concept by applying it to the NREL 6-junction solar cell. We first fit our non-radiative model to the cell. Then, we simulate the performance of the cell in orbit, but using AM1.5G rather than AM0, since that is the spectrum that the cell was designed for and since it makes no difference for the purposes of this exercise. We find that the 6J cell has an operating efficiency of 34.6\% which is lower than its nominal efficiency of 39.2\% since the operating temperature, 337~K, is greater than 298~K. Figure \ref{fig:figure5}a shows a schematic of the cell, labeled "A", along with its nominal and operating efficiencies. 

We then remove the lowest junction, colored red, and again solve for the nominal and operating efficiencies. Compared to the 6J cell, the nominal efficiency decreases by 1.4\%, which is expected. However, under operation in space, the 5J cell actuall outperforms the 6J cell, despite rejecting more of the solar spectrum. 

Figure \ref{fig:figure5}b plots the efficiency versus temperature of each cell. The vertical dotted line demarcates 298~K and the corresponding nominal efficiencies. Points A and B correspond to cells A and B in Figure \ref{fig:figure5}a and demark the operating temperatures and efficiencies. Two things are immediately apparent: first, the 6-junction cell has a steeper temperature dependence compared to the 5-junction cell, and second, the 5J cell operates at a lower temperature and for that reason has a higher efficiency. The 6J cell actually has a higher efficiency at the operating temperature of the 5J cell, however, it is unable to operate at that temperature due to the excess heat generated in attempting to convert the solar infrared tail. Note that in our calculation, we include thermal emission from the earth, approximating it as a blackbody at 255~K, using the effective radiating factor to determine how much of this thermal energy is absorbed.

\section{Discussion}

This work examines the trade-off between harvesting the solar infrared tail and minimizing heat generation in space-based photovoltaics. Longer-wavelength photons generate more waste heat per watt of electricity, and this elevates the operating temperature, which lowers device efficiency, and overall reduces power output. Thus the optimum $\lambda_{cutoff}$ for a photovoltaic device operating with limited heat rejection capability in orbit is different from the room-temperature optimum. 

The effective radiating factor, $A\epsilon/A_{0}$, is the key element that modulates the trade-off between generating waste heat from harvesting the solar tail and the decrease in efficiency due to the rise in operating temperature. Raising $A\epsilon/A_{0}$ increases thermal emission at a given temperature, permitting deeper IR harvesting. For example, for a 4-junction radiative-limit solar cell with $A\epsilon/A_{0} = 1.0$, $\lambda_{cutoff} = 1525~nm$ ($T_{op} = 308~K$) compared to the constant, room-temperature optimum at $\lambda_{cutoff} = 1800~nm$. Increasing $A\epsilon/A_{0}$ to $1.5$ extends the cutoff to $1670~nm$ ($T_{op}=279~K$). Note that $\lambda_{cutoff}$ for $A\epsilon/A_{0} = 1.5$ is still shorter than the room-temperature cutoff despite the operating temperature being below room temperature. This is because the device produces more power from operating at a lower temperature than it would from including more of the solar infrared tail but operating at an elevated temperature. In contrast, the room-temperature optimum cutoff was determined independently of waste heat generation since the device is held at room temperature.

A case study comparing an NREL six‐junction (6J) cell to a truncated five‐junction (5J) variant under AM1.5G illustrates these principles. The 6J cell’s nominal 39.2\% efficiency at 298\,K drops to 34.1\% in orbit ($T_{op} = 349~K$). After removing the bottom junction ($\lambda_{cutoff}$ from 1800\,nm to 1300\,nm) yields a 5J cell with 37.6\% nominal efficiency, which has a higher efficiency in orbit of 34.6\% efficiency ($T_{op} = 337~K$). Truncating the IR tail thus increases net output despite lower room‐temperature performance.

These findings imply that spacecraft photovoltaic arrays must be designed with thermal constraints as a primary criterion. Nominal efficiency metrics that neglect in-orbit temperature effects are misleading: a lower room-temperature efficiency cell with a truncated IR cutoff may outperform a higher room-temperature efficiency design in space. Device designers should prioritize reducing non-radiative and resistive loss mechanisms rather than extending the absorption band. Increasing junction count can still be advantageous, but it should serve the reduction of thermalization losses within a constrained absorption band rather than than the conversion of longer-wavelength photons. Beyond the increase in efficiency from operating at a lower temperature, devices that reject the solar tail may experience other benefits such as being more robust against solar flare events. In addition, lower operating temperature may lead to longer device lifetimes.

\section{Methods}
\subsection{Computational modeling methods (detailed-balance efficiency, thermal modeling equations)}

\subsubsection{Radiative-limit Model}

We built a model that follows the detailed balance limit as first laid out by Shockley and Queisser.\cite{shockley_detailed_1961}. 

The radiative recombination rate can be determined as a function of voltage for there, we solve for the radiative recombination rate:
\begin{align}
    R_{\mathrm{rad}}(V) 
    &= \int_{0}^{\infty}
    \frac{4\pi\,c}{\lambda^{4}}
    \;\frac{1}{\exp\!\Bigl(\frac{h\,c}{k_{B}\,T\,\lambda}
    \;-\;\frac{q}{k_{B}\,T}\,\bigl[V + J\,R_{s}\bigr]\Bigr)\;-\;1}
    \;d\lambda
\end{align}
This is sufficient to determine the performance of a radiative recombination limited cell. However, since we aim to eventually add other recombination modes, we must find the radiative recombination minority carrier lifetime and use this value to determine the excess carrier concentration. 
\begin{align}
    \tau_{rad} = \frac{1}{B_{rad}*N}
\end{align}
Where $B_{rad}$ is the radiative recombination coefficient (a material dependent quantity) and and $N$ is the dopant density; for a solar cell with a p-type base, $N$ is the density of acceptors, and $\tau_{rad}$ gives the electron lifetime. We choose $N_A = 1e16$ for all calculations. For $B_{rad}$ we perform a linear fit of the radiative recombination coefficient versus bandgap for a handful of III-V semiconductors.\cite{Keyes1994,Bach2019,IoffeGaN,IoffeGaSb,IoffeInAs,IoffeInSb}.
\begin{align}
    B_{rad} = 4*10^{-11} e^{1.5662 * Eg_eV}
\end{align}
We then calculate the excess carrier concentration as:
\begin{align}
    \Delta n = R_{rad}\tau_{rad}
\end{align}
Later on, other recombination modes can be included by determining the carrier lifetimes and combining them:
\begin{align}
     \frac{1}{\tau_{tot}} =  \frac{1}{\tau_1} +  \frac{1}{\tau_2} + \cdots + \frac{1}{\tau_n}
\end{align}
And then the combined recombination can be found as:
\begin{align}
    R_{tot} = \frac{\Delta n}{\tau_{tot}}
\end{align}
From there, the dark saturation current is found: 
\begin{align}
    J_{dark} = qR \\
    J_0 = \frac{J_{dark}}{e^\frac{qV}{k_BT}-1}
\end{align}
Using a voltage step of 1~mV, we increment from V=0 and repeat the calculation above until $J_0$ stabilizes. We then use this value of $J_0$ in our diode equation:
\begin{align}
    J_{\mathrm{new}} 
    &= J_{\mathrm{sc}} 
    - J_{0}\,\biggl(\exp\!\Bigl(\frac{q\,\bigl(V + J\,R_{s}\bigr)}{n\,k_{B}\,T}\Bigr) - 1\biggr)
    - \frac{V + J\,R_{s}}{R_{\mathrm{sh}}}
\end{align}
We calculate the JV curve for each junction in a cell, and find $V_{mp}$ and $J_{mp}$. For cells where only a cutoff wavelength is provided, we find the current-matched junction values as follows. We first split the conversion band photon flux equally between the junctions and solve for $J_{mp}$. We then determine a weighting factor for each junction:
\begin{align}
    w = J_{mp}^{avg}/J_{mp}^i;
\end{align}
where $J_{mp}^{avg}$ is the average short-circuit current density of all the junctions and $J_{mp}^{i}$ is the short-circuit current density of the $i$th junction. We then adjust the cutoffs so that each junction has $w$ times more photon flux. We repeat this process until the junctions are current matched.

\subsubsection{Non-Radiative Recombination Model}
We seek to find a model that approximates non-radiative recombination as a function of temperature and bandgap. We begin with a rudimentary model for Shockley-Read-Hall recombination, which assumes that all traps are mid-gap, and results in the following expression for SRH lifetime \cite{jenny_a_nelson_physics_2003}:

\begin{align}
  \tau_{SRH} = \frac{1}{v_t\,\sigma\,N_t}
\end{align}
where $\sigma$ is the capture cross section [$m^2$] and $N_t$ is the trap density [$m^{-3}$]. 

Work in deep level trap spectroscopy, as well as quantum mechanical modeling, reveals that many traps show Arrhenius-type behavior, causing the capture cross section to be a function of temperature: \cite{henry_nonradiative_1977,makram-ebeid_quantum_1982,meneghini_deep-level_2017}. 

\begin{align}
    \sigma = \sigma_0\,exp{\frac{-E_b}{k_BT}} 
\end{align}

Where $E_b$ is the trap activation barrier, $\sigma_0$ is the high-temperature capture cross section. 

Each trap has its own parameters, varying widely by trap species and host material. In addition, the activation barrier of a mid-gap trap is \textit{not} equal to half of the bandgap. This is discussed at length in the following reference.\cite{peter_blood_electrical_1992}. In short, the bandgap is a quantity of Gibbs free energy whereas the activation energy is an enthalpy term. Nevertheless, the general trend is that the activation energy scales with bandgap. We define our activation barrier as:
\begin{align}
    E_b = \alpha\,(E_g[eV]-1.41)-E_{b,0}
\end{align}
where $E_{b,0}$ is a reference mid-gap trap activation energy for GaAs around 0.1~eV, $E_g$ is the bandgap [eV] of the material to be modeled, and $\alpha$ is a model parameter that tunes the bandgap-dependence of the trap activation energy. For non-zero values of alpha, smaller bandgap materials have smaller trap activation energies and thus higher rates of recombination.

The concentration of trap states also exhibits some trends with bandgap. A plot of atomic density versus bandgap energy reveals that atomic density increases with bandgap following an exponential behavior. Thus for a constant defect percent, the density of trap states increases exponentially as well. Furthermore, more species behave as mid-gap traps in wider-bandgap materials, simply because the larger energy gap provides more “space” for defect levels to fall well away from either band edge. Thus, we describe our trap density as:
\begin{align}
    N_t = N_{t,0}\,exp{(f\,E_g[eV]})
\end{align}
where $f$ dictates the bandgap-dependence of the mid-gap trap concentration.

The product of these two terms gives a linear trap density in units of inverse length. We combine the two terms to reduce the number of model parameters that must be fit. The two pre-exponential factors become one parameter, $N_{t,lin,0}$, and the the exponential terms combine to become:
\begin{align}
    N_{t,lin} = N_{t,lin,0}\,exp{(\frac{-(\alpha\,(E_g-1.41)+E_{b,0})}{k_B\,T}+f\,E_g)}
\end{align}
where $E_g$ is the bandgap of the material in question, measured in [eV], and $N_{t,lin,0}$, $\alpha$, $E_{b,0}$, and $f$ are model parameters which must be fit to experimental data. 

From there, the non-radiative lifetime is calculated as:
\begin{align}
  \tau_{SRH} = \frac{1}{v_t\,N_{t,lin}}
\end{align}

We sweep the four parameters and find the combination that minimizes the error in calculated open-circuit voltage and the temperature dependence of the open-circuit voltage for a three-junction GaInP, GaAs, GaInNAsSb tandem cell.\cite{aho_temperature_2015}. 

\begin{table}[ht]
\centering
\begin{tabular}{|c|c|}
\hline
\textbf{Parameter} & \textbf{Value} \\ \hline
$\alpha$ & 0.23 \\ \hline
$E_{b,0}$ & 0.11 \\ \hline
$f$ & 5.5 \\ \hline
$N_{t,lin,0}$ & 14.5 \\ \hline
\end{tabular}
\caption{Non-Radiative Model Linear Trap Density Parameter Values}
\end{table}

\begin{table}[ht]
\centering
\begin{tabular}{|c|c|c|c|}
\hline
\textbf{Parameter} & \textbf{Model} & \textbf{Paper} & \textbf{Units} \\ \hline
Voc (V) & 1.38, 0.88, 0.41 & 1.31, 0.93, 0.35 & V \\ \hline
Entire Cell (Sum) & 2.67 & 2.59 & V \\ \hline
Voc Coefficients (mV/K) & -2.40, -2.25, -2.04 & -2.50, -2.20, -2.00 & mV/K \\ \hline
Entire Cell (Sum) & -6.69 & -6.70 & mV/K \\ \hline
\end{tabular}
\caption{Voc and Voc Coefficients Comparison: Model vs. Paper}
\end{table}

\section*{Acknowledgements}

This material was based upon work supported by the National Science Foundation under grant no. ECCS-2146577

\bibliographystyle{apsrev4-2}   
\bibliography{main}

\end{document}